\documentclass{article}

\usepackage{graphicx}
\usepackage{graphpap}
\usepackage{epsf}
\usepackage[utf8]{inputenc}
\usepackage[english]{babel}
\newcommand{\aap}{{ A\&A~\/}}

\newcommand{\apj}{{ ApJ~\/}}
\newcommand{\apjl}{{ ApJL~\/}}

\newcommand{\mnras}{{ MNRAS~\/}}

\newcommand{\nat}{{ Nature~\/}}

\newcommand{\sovast}{{ Sov.~Ast.~\/}}

\newbox\grsign \setbox\grsign=\hbox{$>$} \newdimen\grdimen
\grdimen=\ht\grsign
\newbox\simlessbox \newbox\simgreatbox \newbox\simpropbox
\setbox\simgreatbox=\hbox{\raise.5ex\hbox{$>$}\llap
     {\lower.5ex\hbox{$\sim$}}}\ht1=\grdimen\dp1=0pt
\setbox\simlessbox=\hbox{\raise.5ex\hbox{$<$}\llap
     {\lower.5ex\hbox{$\sim$}}}\ht2=\grdimen\dp2=0pt
\setbox\simpropbox=\hbox{\raise.5ex\hbox{$\propto$}\llap
     {\lower.5ex\hbox{$\sim$}}}\ht2=\grdimen\dp2=0pt

\begin{document}

\title{Direct Determination of Hubble Parameter Using Type IIn Supernovae}



\author{
  S.\,Blinnikov$^{1,2,3}$\footnote{sergei.blinnikov@itep.ru},
  M.\,Potashov$^{1,2}$\footnote{marat.potashov@gmail.com},
  P.\,Baklanov$^{1,2}$\footnote{baklanovp@gmail.com},
  and A.\,Dolgov$^{1,2,4}$\footnote{dolgov@fe.infn.i}\\
  $^{1}$Institute for Theoretical and Experimental Physics (ITEP), Moscow 117218, Russia\\
  $^{2}$Novosibirsk State University, Novosibirsk 630090, Russia\\
  $^{3}$Sternberg Astronomical Institute, Moscow State University, Moscow 119992, Russia\\
$^{4}$University of Ferrara and INFN, Ferrara 44100, Italy
}

\date{22 June 2012}



\maketitle



\abstract{
 We introduce a novel approach, a Dense Shell Method (DSM), for measuring
  distances for cosmology.
  It is based on original Baade idea to relate absolute difference
  of photospheric radii with photospheric velocity.
  We demonstrate that this idea works:
  the new method does not rely on the Cosmic Distance Ladder
  and gives satisfactory results
  for the most luminous Type IIn Supernovae.
  This allows one to make them good primary distance indicators for cosmology.
  Fixing correction factors for illustration, we obtain with this method
  the median distance of $\approx 68^{+19}_{-15}$(68\%CL)~Mpc to SN~2006gy
  and median Hubble parameter $79^{+23}_{-17}$(68\%CL)~km/s/Mpc.
  }


\section{Introduction}

Supernovae are among the most luminous phenomena in the Universe, and they
can serve as cosmological distance indicators.
In some cases one can use a standard candle method.
Nobel prize 2011 in physics is given ``for the discovery of the accelerating expansion of the
Universe through observations of distant supernovae''.
Actually, Type Ia supernovae have been used for this.

Although SNe~Ia are not uniform in luminosity, they can be standardised.
The standardisation is based on statistical correlations found for nearby events
\cite{Pskovsky77, Phillips93}.
Thus they are \emph{secondary} distance indicators,
see reviews, e.g. \cite{Leibundgut01,Phillips05}.

Type II supernovae, on the other hand, have a much larger variance in luminosity
and therefore cannot provide an accurate distance by photometry alone.
Nevertheless, their
great advantage is the possibility of direct measurement of distance, e.g.
by Expanding Photosphere Method (EPM) \cite{KK1974} when applied to SNe~IIP.
The development of EPM is the spectral-fitting expanding atmosphere method
(SEAM) \cite{BaronSEAM}.
Thus, Type II supernovae are interesting because
there are ways to make them \emph{primary} distance indicators.
A standard candle assumption and its calibration
is not needed for direct methods.
Applications of SNe~IIP in cosmography do not depend upon the steps of Cosmic
Distance Ladder avoiding their systematic and statistical errors.

Due to absolute weakness of SNe~IIP they cannot be used at large cosmological distances.
In this paper we introduce a novel approach to measuring distances for
cosmology with the help of the most luminous Type IIn Supernovae.
The method is based on the formation of an expanding dense shell in SN~IIn and allows one to find
a linear size of a supernova shell in absolute units and distance to it.
This Dense Shell Method (DSM) is
partly based on ideas introduced in EPM and SEAM,
and partly in Expanding Shock Front Method (ESM) \cite{Bartel2007} used for SNR~1993J.

\section{Classical Baade-Wesselink and Kirshner-Kwan methods}

All researches using EPM for supernovae cite  papers
\cite{Baade1926} and \cite{Wesselink1946}.
Actually, EPM introduced by \cite{KK1974} differs from the classical Baade-Wesselink (BW) method.

Here we repeat briefly the steps of BW approach which we apply in our new method.

Measuring colour and flux at two different times, $t_1$
and $t_2$, one finds the ratio of the star's radii,
$R_2/R_1$ (the same can be found from interferometry).
Using weak lines which are believed to be formed near the photosphere
one can measure, in principle, the photospheric speed $v_{\rm ph} = d R_{\rm ph}/dt$.
Then $\int_{t_1}^{t_2} v_{\rm ph} dt$
would give $\Delta R_{\rm ph} = R_2 - R_1 $.
Knowing $R_2/R_1$ and $R_2 - R_1$, it is easy to solve for the radii.
The ratio of fluxes gives the distance $D$:
\begin{equation}
  D=R_{\rm ph}\sqrt{F_\nu({\rm model}) \over F_\nu({\rm observed})} \; .
  \label{dSeam}
\end{equation}
Actually, finding the distance by equation~(\ref{dSeam}) with
$F_\nu({\rm model})$ is equivalent to Spectral-fitting Expanding Atmosphere Method (SEAM) \cite{BaronSEAM}.
The original BW method is based on a simplifying assumption of a diluted supernova blackbody spectrum,
\begin{equation}
  F_\nu({\rm model})=\pi \zeta^2_\nu B_\nu(T_c) \; .
  \label{flux}
\end{equation}
Here the relation of a true photospheric intensity with blackbody brightness
$B_\nu(T_c)$ is accounted for by a correction factor $\zeta_\nu$.
This factor is often called the \emph{dilution} factor (a ratio of a thermalisation radius to $R_{\rm ph}$).
Thus, the distance is:
\begin{equation}
  D=\zeta_\nu R_{\rm ph}\sqrt{\pi B_\nu(T_c) \over F_\nu({\rm observed})}.
\label{dzeta}
\end{equation}
Apart from the correction for the dilution one needs also a correction for limb darkening,
or brightening, for the ratio of pulsation velocity
to the radial velocity accounted for by projection factor $p$.
See, e.g. \cite{Gautschy1987,Sabbey1995,Storm2011b,RastorguevDambis2011,Neilson2012} and references
therein for a discussion of those non-trivial questions on the projection factor and other problems
related to BW method in Cepheids.

In reality, one can measure directly only the matter velocity $v_{\rm m}$ on the photospheric level.
The assumption $v_{\rm m} = v_{\rm ph}$
does not work (as a rule) in exploding stars.
Even for Cepheids this was questioned already by \cite{Whitney1955}.
Velocity of matter at the photosphere of a supernova is not at all
$d R_{\rm ph}/dt$.
The $v_{\rm ph}$ and $v_{\rm m}$ may even have different signs.
That is why the main idea of EPM for SNe is different from BW.

Kirshner and Kwan \cite{KK1974} also used the weak lines
to measure the matter velocity on photospheric level, $v_{\rm m}$,
but they never put $v_{\rm m}$ equal to $d R_{\rm ph}/dt$.
That is why the EPM for supernovae
should be called not the Baade-Wesselink method, but more properly the
Kirshner-Kwan method (KK).
They determine the photospheric radius from the relation
\begin{equation}
  R_{\rm ph}=v_{\rm ph}(t - t_0)\; ,
  \label{freeExp}
\end{equation}
where $t_0$ is the constant close to the explosion epoch.
This relation is based on the assumption of free expansion.
If $R_{\rm ph}$ is obtained, the distance $D$ to the supernova is found from equation~(\ref{dzeta}).

\section{Direct distance determination by the new method}

Let us introduce briefly the essence of the new Dense Shell Method (DSM).

Supernovae of type IIn, contrary to SNe~IIP, do not enter the coasting free expansion phase and
both EPM \cite{KK1974} and SEAM \cite{BaronSEAM} are not directly applicable.
Nevertheless, in SN~IIn case we can use slightly modified classical BW method.
There is a lot of dense matter around the supernova and the shock cannot
break out through the circumstellar shell for months or even years.
Yet, it is clear from our results on SNe~IIn \cite{ChugaiEa04,WooBliHeg2007}
that all matter behind the shock is cooled down by radiation and compressed into a cold dense shell.
One has to measure \emph{wide} emission components of lines and determine velocity of matter
in the dense shell $v_{\rm ds}$ (with highest possible accuracy).
Since forward and reverse shocks are both glued together in this shell the photosphere
moves with the matter as well.
In the dense shell, $v_{\rm ph}$ is exactly equal to the rate of change of $R_{\rm ph}$,
i.e. $v_{\rm ph}=d R_{\rm ph}/dt=v_{\rm ds}$ -- and this can be measured.
Everything looks as Baade suggested already in 1920s!

First, we formulate the DSM for broad-band flux $F$ and integrated correction factor $\zeta$.
The observed flux is $F=\zeta^2 R_{\rm ph}^2 \pi B(T) /D^2$, where $B(T)$ is the blackbody intensity
and $D$ is the photometric distance.
Then $\sqrt{F}=\zeta R_{\rm ph} \sqrt{\pi B(T)} /D$.
The effective blackbody temperature $T$ is measurable, as well as $dR_{\rm ph}$ and $d\sqrt{F}$,
while $D$ does not change.

Hence, if $T$ and $\zeta$ are almost constant between the two measurements, we have
\begin{equation}
  d\sqrt{F}=\zeta dR_{\rm ph} \sqrt{\pi B(T)} /D ,
  \label{dFdist}
\end{equation}
and
\begin{equation}
  D=\zeta dR_{\rm ph} \sqrt{\pi B(T)} / d\sqrt{F} .
  \label{dist}
\end{equation}

Thus, measuring $d\sqrt{F}$, $dR_{\rm ph}$ and $T$, and calculating $\zeta$ from a model,
we find the distance $D$ by a direct method
without any ladder of cosmological distances.

One may limit oneself with this ``two-point'' method for quick evaluation of the distance

However, this quick estimate may result in a large error when fluxes are close to each other
and $d\sqrt{F}$ is small in denominator of equation~(\ref{dist}).
For a more accurate treatment one has to develop the new robust technique for monochromatic
or broad-band fluxes, correction factors, and variable colour temperature $T(t,\nu)$.

If temperature $T$ changes significantly with time $t$ and frequency $\nu$,
we have to rely on the evolution of $R_{\rm ph}$,
which is controlled by the changes of radii $dR_{\rm ph}(t)$ taken from observations.
Moreover, we have also to use a model to calculate a correction factor $\zeta_\nu$
and theoretical flux $F_\nu$.

Assume that the observations are sufficiently frequent to allow us to measure
the increments in radius $dR_{\rm ph}=v_{\rm ph} dt$ for a number of points,
where $dt$ is a difference of time of the successive observations.

Let the initial radius (unknown to us) is $R_0$, and
$R_i \equiv R_0+\Delta R_i$ for $i=1,2,3, \ldots$, where $\Delta R_i$
is already known from the $dR$ integration over time.

Then
\begin{equation}
  \zeta_{\nu i}^2 (R_0+\Delta R_{i})^{2} \pi B_{\nu}(T_{c \nu i}) = 10^{0.4A_\nu} D^2 F_{\nu i}
  \label{distAv}
\end{equation}
or, by taking the root,
  \begin{equation}
  \zeta_{\nu i} (R_0+\Delta R_{i}) \sqrt{\pi B_{\nu}(T_{c \nu i})} = 10^{0.2A_\nu} D \sqrt{F_{\nu i}} .
  \label{sqrtDistAv}
\end{equation}

Here $A_\nu$ is the extinction in stellar magnitudes for the frequency $\nu$.
A good model gives us a set of the $\zeta_{\nu i}, \; T_{c \nu i}$ for all observational points.
From the measured $F_{\nu i}, \; \Delta R_{i}$ we can find $R_0$ and the combination
$a_s \equiv 10^{0.4A_s} D^2$
(where instead of $\nu$ we use index $s$ labelling one of the broad-band filters)
by the least squares method.

To find the distance $D$ we need to know $A_s$ from the astronomical observations,
or we can try to get it from equation~(\ref{sqrtDistAv}) written for different spectral filters.

Knowing $R_0$ we obtain the set of equations:
$$
  10^{0.4A_s} D^2 = a_s ,
$$
This gives us difference $A_{s1} - A_{s2}$, and with the help of, e.g., \cite{Cardelli1989} law
one may find $A_s$.

Actually, we have a set of different trial models with different trial distances.
All unsuitable models (which do not reproduce the shapes of time-dependence of fluxes and colours
with reasonable accuracy when scaled to a proper distance) are discarded.
``Suitable'' means that they reproduce the observed values
of velocity $v$, temperature $T$, and circumstellar envelope density $\rho$.
Scaling means that they have different radii of the dense shell for any given time.
Hence they correspond to different distances to SN.

The high luminosity of type IIn supernovae is explained by inflowing matter merging with the dense shell
in highly radiative shocks.
From the continuity of mass we find:
\begin{equation}
  \frac{v_S}{v_S - v_1} = \frac{\rho_1}{\rho_0} \gg 1 \,
  \label{vShock}
\end{equation}
where $v_S$ is the shock velocity,
$\rho_0$ is the density ahead of the front,
and $v_1$, $\rho_1$ are velocity and density behind the front respectively,
see Fig.~16 in \cite{ChugaiEa04} and Figs.~S10,~S11 in \cite{WooBliHeg2007}.
Due to the extremely high density contrast we have $v_1 \approx v_S$ (hence merging of forward and
reverse shocks).
This pattern of the flow is obtained when all kinetic energy of inflowing matter
in the rest frame of the shock front is radiated away:
\begin{equation}
  F_{\rm rad} = \frac{\rho_0 v_S^3}{2} = \frac{\rho_0 v_1^3}{2}
  \label{shockRadiated}
\end{equation}
Thus the observed broad line components permit to determine $v_S$, and the location of
of the radiation flux creation (and hence the thermalisation radius).
As found in our computations, the latter is close to the photosphere.
This means that the values of $\zeta$ do not vary too much from model to model
and do not strongly depend on the photospheric radius.
That is, all our ``suitable'' models will give us a correct distances $D$
from solving the system of equations~(\ref{sqrtDistAv})
\emph{in one step} despite of the differences in the initially assumed distances.

Other results are obtained by applying EPM of \cite{KK1974} (KK method) to classical type for SNe~IIP.
First \cite{Eastman1996, Schmidt1994} found that $\zeta$ depends primarily on temperature $T$
and can be applied for different photospheric radii.
But this proved to be wrong!
E.g. \cite{Dessart2006} have corrected this statement in case of SN~1999em, where they have obtained
systematically larger correction factors than \cite{Eastman1996, Schmidt1994}.
Our models for SN~1999em \cite{Baklanov2005} support this conclusion quite independently \cite{sebColl05}.
The correction factor for SNe~IIP is more sensitive to the model photospheric radius, than
in our SNe~IIn models with their dense shell: larger is the radius of the SNe~IIP higher is the scattering
in its atmosphere, and hence larger is $R_{\rm ph}$ relative to the thermalisation radius.
Thus, $\zeta$ depends not only on $T$, but also on $R_{\rm ph}$ of supernova of type IIP.
In our case, we have a formation of a dense shell and $\zeta$ is virtually independent of its
radius.
Therefore, in EPM or better to say in KK method one has to \emph{iterate} a set of models
with system~(\ref{dzeta}), (\ref{freeExp}) to obtain self-consistent values of $\zeta$ and distance $D$.

%
%


\section{Distance and Hubble's parameter}

\begin{figure}
  \centering
  \includegraphics[width=0.45\textwidth]{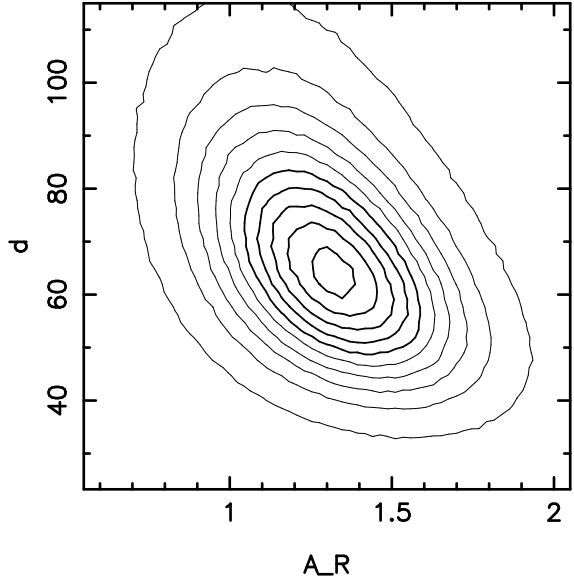}
  \caption
  {
    Monte Carlo resampling simulation of the distance $D$ to SN~2006gy by DSM method.
    The isocontours of probability distribution function (pdf) are shown with equal step in pdf.
    The observations from \cite{Smith2007, Smith2010} have been used for six different time points
    from Table~\ref{table:1} (the first point at $t=36.03$~d is discarded).
    \label{pld06gy}
  }
\end{figure}

For illustration we have taken observational data for SN~2006gy from \cite{Smith2007, Smith2010}.
Unfortunately, the number of epochs for measured temperature is less then the number of observations for fluxes.
We have collected the suitable data points in Table~\ref{table:1} with interpolation in temperature.

We have adopted $v = 5200 \pm 320$~km/s from \cite{Smith2010}. That is the value corresponding to the rising part of the light curve
when $\zeta \approx 1$ and the shell does not fragment.

There are several suggested values for the extinction $A_R$ \cite{Agnoletto2009, Smith2007}.
We have taken $A_R=1.3 \pm 0.25$ mag following \cite{Agnoletto2009}, see discussion in \cite{Moriya2012}.

To estimate the confidence intervals of the distance and $H_0$ we have done a resampling Monte Carlo (MC)
simulation based on these data.
We resampled the values of $T$, of the stellar magnitude $m_R$ in standard filter $R$ \cite{Bessell2005},
the reddening $A_R$, and velocity $v$ each with normal distribution
having standard deviations $\sigma$ from Table~\ref{table:1}.

For obtaining the confidence intervals for the mean and median it was sufficient to do $10^5$ MC tests.
The plot in Fig.~\ref{pld06gy} is built with $10^7$ samples to obtain a better statistics near the top
of the distribution.

Using all 7 points in Table~\ref{table:1} we have obtained the
mean distance $D \approx 63.5$~Mpc, and median $D \approx 62.6$~Mpc
with $68\%$ confidence interval $(-16, +19)$~Mpc.

This simulation used the correction factor $\zeta=1$,
which is close to the values of $\zeta$ with accuracy about $\sim 10$~\% found in our
radiation hydro models \cite{WooBliHeg2007} for the growing part of the light curve.
Of course, an accurate modelling requires building a hydrodynamical model not only for the light curve
but also for spectral line profiles with account of dilution and projection effects as is being done
for recent Cepheid models.

We see that $\sigma_T$ of the first point at $t=36.03$~d in Table~\ref{table:1} is very large.
If we artificially reduce it by a factor of 10 we find
the mean distance $D \approx 73.4$~Mpc, and median $D \approx 72.3$~Mpc
with the $68\%$ confidence interval $(-15,+18)$~Mpc.

If we discard this point completely we get
the mean distance $D \approx 70.3$~Mpc, and median $D \approx 68.2$~Mpc
with $68\%$ confidence interval $(-15,+19)$~Mpc.
Those experiments show that the results are rather robust given the level of accuracy
of data and models.

\begin{table}
\caption{Observations of SN~2006gy} 
\label{table:1}                     
\centering                          
\begin{tabular}{c c c c c c}        
\hline\hline                        
  time, d  & $T, 10^3$~K & err.($T$) & $m_R$ & err.($m_R$) \\ 
\hline                              
  36.03 &12  & 3  &14.72 & .03 \\  
  40.95 &12  & .8 &14.62 & .03 \\
  47.97 &12  & .8 &14.42 & .03 \\
  59.92 &12  & .8 &14.27 & .03 \\
  71.0, &11  & .7 &14.22 & .03 \\
  82.92 & 9  & .8 &14.28 & .03 \\
  94.88 & 8.8& .4 &14.49 & .03 \\
\hline                              
\end{tabular}
\end{table}

This value of distance $D$ is in good agreement with a generally accepted value 71~Mpc, see Fig.~\ref{pld06gy}.
The largest thick-line contour in Fig.~\ref{pld06gy} is about one standard deviation.
The error of our value is quite high mainly because of the uncertainty of temperature $T$ and reddening $A_R$.
Nevertheless, even this accuracy is enough to make quite implausible the suggestion \cite{Arp2007} to put
SN~2006gy much closer to us, around 10~Mpc.
The supernova itself ``tells'' us, that its distance is an order of magnitude larger than 10~Mpc.

Using the redshift $z=0.0179$ for the galaxy NGC~1260, where SN~2006gy has exploded,
we obtain the Hubble parameter.
We do not use directly the formula
\begin{equation}
  H_0 = \frac{c z}{D}
\end{equation}
since, e.g., median($H_0$) is not equal $cz/\mbox{median}(D) $.
So, our values of $H_0$ are computed as a result of MC for each individual $D$.

Thus, with all data of Table~\ref{table:1} we get
the mean $H_0 \approx 95.2$~km/s/Mpc and the median $H_0 \approx 85.7$~km/s/Mpc
with the $68\%$ confidence interval $(-20, +29)$~km/s/Mpc.

If we reduce the error of the first point by a factor of 10 we find
the mean $H_0 \approx 76.5$~km/s/Mpc and
the median $H_0 \approx 74.2$~km/s/Mpc
with $68\%$ confidence interval $(-15, +19)$~km/s/Mpc.

When the first point is discarded, we get
the mean $H_0 \approx 81.5$~km/s/Mpc and
the median $H_0 \approx 78.7$~km/s/Mpc
with $68\%$ confidence interval $(-17, +23)$~km/s/Mpc.

The latter result is the most reliable, so the median for the Hubble parameter is
\begin{equation}
  H_0 \approx 79^{+23}_{-17} \; \mbox{km/s/Mpc} .
\end{equation}

The accuracy is about 30\%, mainly influenced by the error in the temperature and the interstellar extinction \cite{Smith2010}.
Of course, this accuracy of $H_0$ is low compared with the one already achieved by other techniques,
but our value is obtained by the new direct method and does not rely on the Cosmic Distance Ladder.

Statistics of similar objects with more precise reddening can significantly improve the $H_0$ accuracy
in the future.
It is needed to investigate the role of variations of the correction factors in different SN~2006gy models
to check the robustness of our results.
We present here the values for $D$ and $H_0$ only for the illustration of the efficiency of the method.

\section{Conclusions}

Now, we can summarise essential features of the new method, DSM (Dense Shell Method),
for finding cosmological distances with the help of SNe~IIn.
The method is based on the following steps:

\begin{itemize}

  \item
  Measurement of \emph{wide} emission components of lines
  and determination of the velocity at photosphere level
  $v_{\rm m} = v_{\rm ph}$ (with highest possible accuracy).

  \item
  Measurement of \emph{narrow} components of spectral lines
  for estimating properties (density, velocity) of circumstellar envelope.
  One does not need a very high accuracy of measurements and modelling here.

  \item
  Building of a set of \emph{best fitting models} (``suitable'') for broad band photometry and
  speed $v_{\rm ph}$, for a set of trial distances
  satisfying the constraints for the circumstellar envelope found
  from narrow lines.

  \item
  Although the free expansion assumption $v = {r}/{t}$ is not applicable,
  $v_{\rm m}$ now measures a true velocity of the photospheric radius
  (not only the matter flow speed, as in type IIP).

  \item
  Now the original Baade's idea works for measuring
  the radius $R_{\rm ph}$ by integrating $dR_{\rm ph}=v_{\rm ph} dt$
  (of course, with due account of scattering, limb darkening/brightening etc. in a time-dependent modelling).

  \item
  The observed flux then gives the {\it distance} $D$ from the system~(\ref{sqrtDistAv}).

\end{itemize}

The constraining of cosmological parameters and our understanding of
Dark Energy depend strongly on accurate measurements of distances in Universe.
SNe~IIn may be used for cosmology as \emph{primary distance indicators} with the new DSM method.
Application of EPM and SEAM requires crafting a best fitting
hydro model for each individual SN.
This procedure is in principle simpler in DSM.
The case of SN~2006gy shows that the DSM distance agrees well
with other most reliable techniques when the correct model is used,
without the assumption on free expansion which is needed for EPM and SEAM.

\section*{Acknowledgements}

SB and PB are grateful to Wolfgang Hillebrandt  for hospitality at MPA.
SB thanks Ken Nomoto for years of collaboration and support at IPMU, and Norbert Langer, Sung-Chul Yoon,
and Hilding Neilson for valuable discussions at AIfA.
The work is supported partly by the grants
of the Government of the Russian Federation (No 11.G34.31.0047),
by RFBR 10-02-00249, 10-02-01398,
by RF Sci.~Schools 3458.2010.2, 3899.2010.2,
and by a grant IZ73Z0-128180/1 of the Swiss National Science
Foundation (SCOPES).


\end{document}